\documentstyle[aps,preprint,tighten]{revtex}
  
\newcommand{\sn}{{\rm sn}}
\newcommand{\cn}{{\rm cn}}
\newcommand{\dn}{{\rm dn}}

\newcommand{\nc}{{\rm nc}}

\newcommand{\uT}{u_{\Theta}}
\newcommand{\ut}{u_{\theta}}
\newcommand{\vT}{\varphi_{\Theta}}

\begin{document} 
  
  

\title{Semiclassical Series from Path Integrals\footnote{Published
in {\it Trends in Theoretical Physics II} (AIP Conference Proceedings
484), edited by H.\ Falomir, R.\ E.\ Gamboa Sarav\'{\i}, and
F.\ A.\ Schaposnik (American Institute of Physics, Woodbury, 1999)
pp 256--269.}}

\author{C.\ A.\ A.\ de Carvalho\footnote{E-mail: aragao@if.ufrj.br}} 

\address{Instituto de F\'\i sica, Universidade Federal do Rio de Janeiro,  
\\ Cx.\ Postal 68528, CEP 21945-970, Rio de Janeiro, RJ, Brasil} 
 
\author{R.\ M.\ Cavalcanti\footnote{Present address: Instituto de 
F\'\i sica, 
Universidade de S\~ao Paulo, Cx.\ Postal 66318, CEP 05315-970, 
S\~ao Paulo, SP, Brasil. E-mail: rmoritz@fma.if.usp.br}} 
 
\address{Institute for Theoretical Physics, University of California,
\\ Santa Barbara, CA 93106-4030, USA} 
 
 
\maketitle

\begin{abstract}
 
We derive the semiclassical series for the partition function in Quantum 
Statistical Mechanics (QSM) from its path integral representation. Each 
term of the series is obtained explicitly from the (real) minima of
the classical action. The method yields a simple derivation of the
exact result for the harmonic oscillator, and an accurate estimate of 
ground-state energy and specific heat for a single-well quartic
anharmonic oscillator. As QSM can be regarded as finite temperature
field theory at a point, we make use of the field-theoretic language
of Feynman diagrams to illustrate the non-perturbative character of
the series: it contains all powers of $\hbar$ and graphs with any
number of loops; the usual perturbative series corresponds to a subset
of the diagrams of the semiclassical series. We comment on the
application of our results to other potentials, to correlation
functions and to field theories in higher dimensions.   
 
\end{abstract}

\newpage
 
  
\section{Introduction} 
\label{introduction}

Semiclassical series have a long history in Quantum Mechanics which
goes back 
to the early days of the Schr\"odinger equation. Their first terms are 
dictated by classical trajectories, and serve as the initial step of 
iterative procedures which yield all other terms \cite{annals}.

Path integral representations for correlation functions have been developed 
more recently \cite{feynman,schulman,rivers,weiss,kleinert}. Being
sums over 
trajectories, they provide a natural derivation of  semiclassical results 
through the stationary phase method. Indeed, the saddle points of the path 
integrals correspond to classical trajectories which dictate the first 
terms of the series \cite{gutzwiller1,dashen,rajaraman}. 
However, despite the many applications in Quantum Mechanics and Field 
Theory, most discussions which used path integrals never went beyond the 
first term of a semiclassical series. Notable exceptions were the works 
of DeWitt-Morette \cite{morette1} and Mizrahi \cite{mizrahi} in Quantum 
Mechanics.

Semiclassical methods for finite temperature field theories 
\cite{bernard,dolan,weinberg} also remained restricted to derivations of 
the first term of a semiclassical series \cite{ma}, even when the problem 
was reduced to Quantum Statistical Mechanics \cite{harrington,dolankiskis}, 
viewed as field theory at a point (zero spatial dimension). Some references 
resorted to extensions to the complex plane 
\cite{mottola,carlitz,wileyboyanholman} to include complex paths required 
to describe Fourier transformed quantities but, again, those treatments 
were not concerned with obtaining the whole series.

In this talk we will present a systematic path integral 
procedure to generate semiclassical series in Quantum Statistical 
Mechanics. 
It leads to the construction of each term of the series from the 
solution(s) of the classical equations of motion. We will focus our 
attention on the partition function, and use the method of steepest 
descent which, in this case, only requires {\it real} solutions as 
saddle-points \cite{joras}. The restriction of our analysis to 
quantum-mechanical systems (i.e., field theories at a point and at finite 
temperature) will allow us to construct the semiclassical propagator 
needed to generate the terms of the series.

In our contribution to last year's Workshop \cite{laplata}, we had already 
outlined the procedure mentioned in the previous paragraph. However, the 
approach we used then was quantum-mechanical, as opposed to the 
field-theoretic language we shall adopt in the present article. 
Reference \cite{annals} describes both approaches and gives a detailed 
account of the results which will be quoted here.

This article is organized as follows. Section \ref{statistical} presents 
the derivation of the semiclassical series for a generic one-dimensional 
potential of the single-well type in field-theoretic language, which allows 
for a simple connection with the works of references 
\cite{morette1,mizrahi}; the presentation is a natural 
extension of textbook material \cite{zinnjustin}, and profits from the 
clear account of reference \cite{wileyboyanholman}.
Section \ref{oscillators} 
applies our results to the harmonic oscillator and to the single-well 
quartic anharmonic oscillator; for the latter, we compute the 
ground-state 
energy and the specific heat. Section \ref{conclusions} presents our 
conclusions, comments on extensions to double-well type potentials, to 
QSM in higher dimensions and to field theories. 


\section{Quantum Statistical Mechanics}     
\label{statistical}

The partition function for a one-dimensional quantum-mechanical 
system consisting of a particle of mass $m$ in the presence of a 
potential $V(x)$ in equilibrium with a thermal reservoir at 
temperature $\beta^{-1}$ can be written as a path integral:
\begin{equation}
Z(\beta)=\int_{-\infty}^\infty dx_0 \int_{x(0)=x_0}^{x(\beta\hbar)=x_0}
[{\cal D}x(\tau)]\, e^{-S/\hbar},
\end{equation}
\begin{equation}
S[x]=\int_0^{\beta\hbar} d\tau
\left[\frac{1}{2}\,m\left(\frac{dx}{d\tau}\right)^2+V(x)\right].
\end{equation}
For convenience we define the dimensionless quantities 
$q\equiv x/x_N$, $\theta\equiv \omega_N\tau$, 
$\Theta\equiv \beta\hbar\omega_N$, $U(q)\equiv V(x_Nq)/m\omega_N^2x_N^2$
and $g\equiv\hbar/m\omega_Nx_N^2$,
where $\omega_N^{-1}$ and $x_N$
are natural time and length scales of the problem, respectively.
In terms of these quantities we rewrite the partition function as
\begin{equation}
Z(\Theta)=\int_{-\infty}^{\infty}dq_0
\int_{q(0)=q_0}^{q(\Theta)=q_0}[{\cal D}q(\theta)]\, e^{-I/g},
\label{Z}
\end{equation}
\begin{equation}
I[q]=\int_0^\Theta d\theta \left[\frac{1}{2}\,{\dot q}^2
+U(q)\right],
\end{equation}
where the dot denotes differentiation in $\theta$. 

We generate a semiclassical series for $Z(\Theta)$ by: (i) finding the
minima 
$q_c(\theta)$ of the Euclidean action $I$, i.e., the stable classical 
paths that solve the Euler-Lagrange equation of motion, subject to the 
boundary conditions; (ii) expanding the Euclidean action around these 
classical 
paths; (iii) deriving a quadratic semiclassical propagator by neglecting 
terms higher than second order in the expansion; (iv) using that propagator 
to compute higher (than quadratic) order contributions perturbatively.

For the sake of simplicity, we shall restrict our analysis to potentials 
of the single-well type, twice differentiable, and such that $U'(q)=0$
only at the minimum of $U$, which we shall assume to be at the origin 
(see Fig.\ \ref{potential}). This guarantees that, 
given $q_0$ and $\Theta$, there will be a {\em unique} classical path 
satisfying the boundary conditions. Multiple-well potentials force us 
to consider more than one classical path for certain choices of $q_0$ 
and $\Theta$. This phenomenon has been analyzed, for a double-well type 
potential, using the language of catastrophes and bifurcations \cite{bjp}. 
Semiclassical series for the double-well quartic oscillator will be 
presented elsewhere \cite{joras}.

The Euler-Lagrange equation ($U'\equiv dU/dq$)
\begin{equation}
\ddot{q} -U'(q)=0,
\label{euler}
\end{equation}
subject to the boundary conditions $q(0)=q(\Theta)=q_0$, describes the
motion of a particle in the potential {\em minus} $U$. Its first integral is
\begin{equation}
\frac{1}{2}\,{\dot q}^2=U(q)-U(q_t),
\label{motion}
\end{equation}
where $q_t$ denotes the single turning point (since we have an
inverted single well) of the motion, defined implicitly by
\begin{equation}
\Theta=2\int_{q_0}^{q_t} \frac{dq}{v(q,q_t)},
\label{theta}
\end{equation}
where $v(q,q')\equiv {\rm sign}(q'-q)\sqrt{2[U(q)-U(q')]}$; equation 
(\ref{theta}) is a consequence of integrating (\ref{motion}). Thus, for  
a single well, given $q_0$ and $\Theta$, the classical path will go from 
$q_0$, at $\theta=0$, to $q_t=q_t(q_0,\Theta)$, at $\theta=\Theta/2$, and 
return to $q_0$ at $\theta=\Theta$. (Note that 
${\rm sign}(q_t)={\rm sign}(q_0)$.)

The action for this classical path has a simple expression in terms of
its turning point:
\begin{equation}
I[q_c]=\Theta\, U(q_t)+2\int_{q_0}^{q_t}dq \, v(q,q_t),
\label{action}
\end{equation}
where we have used (\ref{motion}). The first term in (\ref{action}) 
corresponds to the high-temperature limit of $Z(\Theta)$, where classical 
paths collapse to a point ($q_t\to q_0$). The last term will be 
negligible for potentials that vary little over a thermal wavelength 
$\lambda=\hbar\sqrt{\beta/m}$. However, by decreasing the temperature 
it will become important and bring in quantum effects.

We now expand the action around the classical path. Letting 
$q(\theta)=q_c(\theta)+\eta(\theta)$, with 
$\eta(0)=\eta(\Theta)=0$, we obtain
\begin{equation}
I[q]=I[q_c]+I_2[\eta]+\delta I[\eta],
\label{fluc}
\end{equation}
where
\begin{equation}
I_2[\eta] \equiv \frac{1}{2}\int_0^\Theta d\theta\,\left\{
\dot\eta^2(\theta)+U''[q_c(\theta)]\,\eta^2(\theta)\right\},
\end{equation}
\begin{equation}
\delta I[\eta] \equiv \int_0^\Theta d\theta\, \delta U(\theta,\eta)
=\sum_{n=3}^\infty \frac{1}{n!}\int_0^\Theta d\theta\,
U^{(n)}[q_c(\theta)]\,\eta^n(\theta).
\label{deltaI}
\end{equation}
Inserting (\ref{fluc}) into (\ref{Z}) and expanding $e^{-\delta I/g}$
in a power series yields
\begin{equation}
Z(\Theta)=\int_{-\infty}^\infty dq_0\,e^{-I[q_c]/g}
\int_{\eta(0)=0}^{\eta(\Theta)=0}
[{\cal D}\eta(\theta)]\, e^{-I_2[\eta]/g}
\sum_{m=0}^\infty \frac{1}{m!}\left(-\frac{\delta I[\eta]}{g}\right)^m.
\label{zpow}
\end{equation}
The summation in (\ref{zpow}) can be written more explicitly as
\begin{equation}
\sum_{m=0}^\infty \frac{1}{m!}\left(-\frac{\delta I[\eta]}{g}\right)^m=1+
\sum_{m=1}^\infty\frac{(-1)^m}{g^m m!}\prod_{j=1}^m\left[\sum_{n_j=3}^\infty
\frac{1}{n_j!}\int_0^\Theta d\theta_j\, U^{(n_j)}[q_c(\theta_j)]
\,\eta^{n_j}(\theta_j)\right].
\label{sumexp}
\end{equation}
As a consequence, one is led to compute integrals of the following type:
\begin{equation}
\langle\eta(\theta_1)\cdots\eta(\theta_k)\rangle\equiv
\int_{\eta(0)=0}^{\eta(\Theta)=0}[{\cal D}\eta(\theta)] 
\,e^{-I_2[0,\Theta;\eta]/g}\,\eta(\theta_1)\cdots\eta(\theta_k).
\label{etaeta}
\end{equation}
Such integrals emerge naturally as functional 
derivatives of the following generating functional:
\begin{equation}
{\cal Z}[J]=
\int_{\eta(0)=0}^{\eta(\Theta)=0}[{\cal D}\eta(\theta)] 
\,e^{-\frac{1}{g}\left\{I_2[0,\Theta;\eta]-\int_{0}^{\Theta}
d\theta\, J(\theta)\,\eta(\theta)\right\}}.
\label{gtil}
\end{equation}
Indeed,
\begin{equation}
\langle\eta(\theta_1)\cdots\eta(\theta_k)\rangle =
g^k\,\frac{\delta^k\,{\cal Z}[J]}
{\delta J(\theta_1)\cdots\delta J(\theta_k)}\Bigg|_{J=0}.
\label{bracket}
\end{equation}

In order to compute ${\cal Z}[J]$, we define
\begin{equation}
\eta(\theta)=\tilde\eta(\theta)+\int_{0}^{\Theta}d\theta'\, 
{\cal G}(\theta,\theta')\,J(\theta'),
\label{tileta}
\end{equation}
where $\tilde\eta(0)=\tilde\eta(\Theta)=0$, and ${\cal G}(\theta,\theta')$
satisfies
\begin{equation}
\left\{-\frac{\partial^2}{\partial\theta^2}+U''[q_c(\theta)]\right\}
{\cal G}(\theta,\theta')=\delta(\theta-\theta'),
\qquad{\cal G}(0,\theta')={\cal G}(\Theta,\theta')=0.
\label{green}
\end{equation}
Inserting (\ref{tileta}) in (\ref{gtil}), and noting that
$[{\cal D}\eta(\theta)]=[{\cal D}\tilde\eta(\theta)]$,
we obtain
\begin{equation}
{\cal Z}[J]=e^{\frac{1}{2g}\int_0^{\Theta}d\theta
\int_0^{\Theta}d\theta'\,J(\theta)\,{\cal G}(\theta,\theta')\,
J(\theta')}\int_{\tilde\eta(0)=0}^{\tilde\eta(\Theta)=0}
[{\cal D}\tilde\eta(\theta)]\,e^{-I_2[0,\Theta;\tilde\eta]/g}
\label{ZJ}
\end{equation}
If we define 
\begin{equation}
G_c(\theta_1,\eta_1;\theta_2,\eta_2)=
\int_{\eta(\theta_1)=\eta_1}^{\eta(\theta_2)=\eta_2} 
[{\cal D}\eta(\theta)]\, e^{-I_2[\theta_1,\theta_2;\eta]/g},
\label{gc}
\end{equation}
\begin{equation}
I_2[\theta_1,\theta_2;\eta]=\frac{1}{2}\int_{\theta_1}^{\theta_2} d\theta\, 
\left\{{\dot \eta}^2 +U''[q_c(\theta)]\, \eta^2\right\},
\label{i2}
\end{equation}
we finally arrive at
\begin{equation}
{\cal Z}[J]=G_c(0,0;\Theta,0)
\,\exp\left[\frac{1}{2g}\int_{0}^{\Theta}d\theta \int_{0}^{\Theta}d\theta' 
J(\theta)\,{\cal G}(\theta,\theta')\, J(\theta')\right].
\end{equation}
Using this result, we can now calculate (\ref{bracket}).
The result is simply
\begin{equation}
\langle\eta(\theta_1)\cdots\eta(\theta_k)\rangle =g^{k/2}\,
G_c(0,0;\Theta,0)\,
\sum_P{\cal G}(\theta_{i_1},\theta_{i_2})\cdots 
{\cal G}(\theta_{i_{k-1}},\theta_{i_k}),
\label{sumP}
\end{equation}
if $k$ is even, and zero otherwise. $\sum_P$ denotes sum over all 
possible pairings of the $\theta_{i_j}$. Inserting this into (\ref{zpow}) 
and (\ref{sumexp}) yields the semiclassical series for $Z(\Theta)$.

We still have to solve Eq.\ (\ref{green}). This can be easily done if one 
notes that, for $\theta\ne\theta'$, it is a homogeneous second-order 
differential equation. Therefore,
${\cal G}(\theta,\theta')$ can be constructed from a linear 
combination of two linearly independent solutions $\eta_a(\theta)$ and 
$\eta_b(\theta)$ of the equation
\begin{equation}
\ddot \eta - U''[q_c(\theta)]\,\eta=0.
\label{extremum}
\end{equation}
Indeed
\begin{equation}
{\cal G}(\theta,\theta')=\left\{
\begin{array}{ll}
a_-\eta_a(\theta)+b_-\eta_b(\theta),& \theta<\theta' \\
a_+\eta_a(\theta)+b_+\eta_b(\theta),& \theta>\theta'.
\end{array}
\right.
\label{calG}
\end{equation}
Continuity imposes
\begin{equation}
{\cal G}(\theta'+\epsilon,\theta')={\cal G}(\theta'-\epsilon,\theta'),
\label{cont}
\end{equation}
whereas (\ref{green}) leads to
\begin{equation}
\frac{\partial}{\partial\theta}\,{\cal G}(\theta,\theta')
\Big|_{\theta=\theta'+\epsilon} 
- \frac{\partial}{\partial\theta}\,{\cal G}(\theta,\theta')
\Big|_{\theta=\theta'-\epsilon}=-1,
\label{deriv}
\end{equation}
with $\epsilon\to 0^+$. (\ref{cont}), (\ref{deriv}) and the
boundary conditions completely determine the coefficients
in (\ref{calG}). The final result is
\begin{equation}
{\cal G}(\theta,\theta')=\frac{\Omega(0,\theta_<)\,
\Omega(\theta_>,\Theta)}{\Omega(0,\Theta)},
\label{wronsk}
\end{equation}
where $\theta_<(\theta_>)\equiv{\rm min(max)}\{\theta,\theta'\}$,
and $\Omega(\theta_1,\theta_2)$ is the function
\begin{equation}
\Omega(\theta,\theta')\equiv \eta_a(\theta)\,\eta_b(\theta')
-\eta_a(\theta')\,\eta_b(\theta).
\label{omega}
\end{equation}
In the Appendix we show that $G_c(\theta_1,\eta_1;\theta_2,\eta_2)$
can also be obtained from the two linearly independent solutions of 
Eq.\ (\ref{extremum}), $\eta_a(\theta)$ and $\eta_b(\theta)$; 
furthermore, we show how to construct those two functions from the 
solution $q_c(\theta)$ of the classical equation of motion. This completes 
the steps needed to write down any term of the series: all that is required 
is $q_c(\theta)$! 
 
 
\section{Quantum Oscillators}
\label{oscillators}

In this section we will apply our construction to the harmonic oscillator 
and to the single-well quartic oscillator. The harmonic case is designed 
to illustrate the compactness of our general formulae, which immediately 
yield the exact answer --- there is no need to compute functional 
determinants 
from eigenvalue problems! The anharmonic case is designed to
illustrate their power --- the first term of the semiclassical series 
for the partition function 
allows us to extract a very good estimate of the ground-state energy and of 
the specific heat.

\subsection{The Harmonic Oscillator}
\label{harmonic}

In this subsection, we study the potential 
\begin{equation}
V(x)=\frac{1}{2}\,m\omega^2x^2.
\label{hvx}
\end{equation}
Choosing $\omega_N=\omega$ and $x_N=\sqrt{\hbar/m\omega}$, and
introducing the dimensionless quantities 
of section \ref{statistical}, we have
$g=1$ and
\begin{equation}
U(q)=\frac{1}{2}\,q^2.
\label{hux}
\end{equation}

Integrating (\ref{motion}) leads to
\begin{equation}
q_c(\theta)=q_t\, \cosh (\theta-\frac{\Theta}{2}),
\label{hclassical}
\end{equation}
The relation between $q_0$ and $q_t$ is obtained by taking
$\theta=\Theta$ in (\ref{hclassical}):
\begin{equation}
q_0=q_c(\Theta)=q_t\, \cosh (\Theta/2) .
\label{hq0-qt}
\end{equation}
The action for the classical path is
\begin{equation}
I[q_c]=q_0^2 \tanh(\Theta/2).
\label{haction}
\end{equation}
Following the Appendix, the functions $\eta_a$ and $\eta_b$ are given by 
\begin{equation}
\eta_a(\theta)=\dot q_c(\theta)=q_t\, \sinh(\theta-\frac{\Theta}{2}),
\label{hqatheta}
\end{equation}
and
\begin{equation}
\eta_b(\theta)=\dot q_c(\theta)\,Q(\theta)=-q_t^{-1}\, 
\cosh(\theta-\frac{\Theta}{2}),
\label{hqbtheta}
\end{equation}
It follows that $\Omega_{ij}=\sinh(\theta_j - \theta_i)$ and
$W_{ij}=\cosh(\theta_j - \theta_i)$, leading to
\begin{eqnarray}
G_c(\theta_1,\eta_1;\theta_2,\eta_2)&=&\frac{1}{\sqrt{2\pi \sinh(\theta_2 
- \theta_1)}}
\nonumber \\
& &\times\exp\left\{-\frac{1}{2\sinh(\theta_2 - \theta_1)}
\left[\cosh(\theta_2 - \theta_1)\,(\eta_2^2+\eta_1^2)
-2\,\eta_1\eta_2\right]\right\}.
\label{hgs}
\end{eqnarray}
Using (\ref{vanvleck}) we obtain $\Delta=2\pi\, \sinh\Theta$. Since the 
problem is quadratic, its exact solution is then given by
\begin{equation}
Z(\Theta)\equiv \int_{-\infty}^\infty dq_0\, e^{-I[q_c]/g} 
\Delta^{-1/2}
\label{hz2.1}
\end{equation}
Inserting (\ref{haction}) and the value of $\Delta$ yields the well-known 
result 
\begin{equation}
Z(\Theta)=
\int_{-\infty}^{\infty}\frac{e^{-q_0^2\,\tanh(\Theta/2)}}
{\sqrt{2\pi \,\sinh\Theta}}\,dq_0=\frac{1}{2\,\sinh(\Theta/2)}.
\label{hz2lim}
\end{equation}


\subsection{The Single-well Quartic Oscillator}
\label{quartic}

In this subsection, we study the potential 
\begin{equation}
V(x)=\frac{1}{2}\,m\omega^2x^2+\frac{1}{4}\,\lambda x^4.
\label{vx}
\end{equation}
Choosing $\omega_N=\omega$ and $x_N=\sqrt{m\omega^2/\lambda}$, and
introducing the dimensionless quantities 
of section \ref{statistical}, we have
$g=\lambda\hbar/m^2\omega^3$ and
\begin{equation}
U(q)=\frac{1}{2}\,q^2+\frac{1}{4}\,q^4.
\label{ux}
\end{equation}

Integrating (\ref{motion}) leads to \cite{grads,byrd}
\begin{equation}
q_c(\theta)=q_t\, \nc (u_{\theta},k),
\label{classical}
\end{equation}
where $\nc (u,k)\equiv 1/\cn  (u,k)$ is one of the Jacobian Elliptic 
functions \cite{grads,byrd,as}, and
\begin{equation}
u_{\theta}=\sqrt{1+q_t^2}\left(\theta-\frac{\Theta}{2}\right),
\qquad k=\sqrt{\frac{2+q_t^2}{2\,(1+q_t^2)}}.
\label{u,k}
\end{equation}
For future use, we note that (\ref{u,k}) can be rewritten as
\begin{equation}
u_{\theta}=\frac{2\theta-\Theta}{2\sqrt{2k^2-1}},
\qquad|q_t|=\sqrt{\frac{2\,(1-k^2)}{2k^2-1}}.
\label{gqt}
\end{equation}
The relation between $q_0$ and $q_t$ is obtained by taking
$\theta=\Theta$ in (\ref{classical}):
\begin{equation}
q_0=q_c(\Theta)=q_t\,\nc\,\uT.
\label{q0-qt}
\end{equation}
(We shall often omit the $k$-dependence in the Jacobian Elliptic
functions.)

The action for the classical path (\ref{classical}) is
\begin{equation}
I[q_c]=\Theta\, U(q_t)+\sqrt{2}\int_{|q_t|}^{|q_0|}dq\,
\sqrt{(q^2+q_t^2+2)(q^2-q_t^2)}.
\end{equation}
Performing the integral (Ref.~\cite{grads}, formula 3.155.6) 
and replacing $q_0$ by the r.h.s.\ of (\ref{q0-qt}), we obtain
\begin{eqnarray}
I[q_c]&=&\Theta\left(\frac{1}{2}\,q_t^2+
\frac{1}{4}\,q_t^4\right)+\frac{4}{3}\left\{-\sqrt{1+q_t^2}
\left[{\rm E}(\vT,k)+\frac{1}{2}\,q_t^2\,\uT\right]\right.
\nonumber \\
& &+\left.\sn\,\uT\left(1+\frac{1}{2}\,q_t^2\,\nc^2\uT
\right)\sqrt{1+\frac{1}{2}\,q_t^2\,(1+\nc^2\uT)}\right\},
\label{iqc}
\end{eqnarray}
where ${\rm E}(\varphi,k)$ denotes the Elliptic Integral of the 
Second Kind and
$\varphi_{\theta}\equiv\arccos[q_c(\theta)/q_0]=\arccos(\cn\,\ut)$.

For the construction of the quadratic semiclassical propagators 
$G_c(\theta,\eta;\theta',\eta')$ and ${\cal G}(\theta,\theta')$ we
shall need
\begin{equation}
\eta_a(\theta)=\dot q_c(\theta)=q_t\,\sqrt{1+q_t^2}\,
\sn\, u_{\theta}\,\dn\, u_{\theta}\,\nc ^2u_{\theta}
\label{qatheta}
\end{equation}
and
\begin{eqnarray}
Q(\theta)&=&q_t^{-2}(1+q_t^2)^{-3/2}\left[\left(1-\frac{1}{k^2}\right)
u_{\theta}+\left(\frac{1}{k^2}-2\right){\rm E}(\varphi_{\theta},k)\right.
\nonumber \\
& &-\left.\frac{\cn\,u_{\theta}\,\dn\,u_{\theta}}{\sn\,u_{\theta}}
+(k^2-1)\,\frac{\cn\,u_{\theta}\,\sn\,u_{\theta}}{\dn\,u_{\theta}}\right].
\label{qtheta}
\end{eqnarray}
We may then obtain $\eta_b(\theta)=\dot q_c(\theta)\,Q(\theta)$ and, thus, 
$\Omega_{12}$ and $W_{12}$ from (\ref{o12}) and (\ref{w12}). Finally, 
use of (\ref{gs}) and (\ref{wronsk}) will yield the desired propagators.

For the series expansion of the partition function, 
we shall need
\begin{equation}
\delta U(\theta,\eta)=q_c(\theta)\,\eta^3+\frac{1}{4}\,\eta^4,
\label{deltaU}
\end{equation}
obtained from (\ref{deltaI}). Therefore, we have to consider not only 
the usual quartic vertex, but an additional time($\theta$)-dependent 
cubic term. This completes the set of ingredients needed to write down 
a semiclassical series for any correlation function. 
In the next subsection, we 
shall concentrate on the first term of the series for 
$Z(\Theta)$, which yields the quadratic approximation.


\subsubsection{The quadratic approximation for $Z(\Theta)$}
\label{quadratic}
        
{}From the knowledge of the classical action and of the Van Vleck 
determinant, we define
\begin{equation}
Z_2(\Theta)\equiv \int_{-\infty}^\infty dq_0\, e^{-I[q_c]/g} 
\Delta^{-1/2}
\label{z2.1}
\end{equation}
as the quadratic approximation to $Z(\Theta)$. To perform the
integral over $q_0$ one must write $I[q_c]$ and $\Delta$ solely
in terms of $q_0$ (and $\Theta$), but except in rare cases
this is not an easy task. Usually, it is much simpler to write
these quantities in terms of $q_t$ [see Eq.\ (\ref{q0-qt})],
and so it is natural to trade $q_0$ for $q_t$ as the 
integration variable in (\ref{z2.1}). This is much simplified by the fact 
that the Jacobian of the map $q_0\to q_t$ is simply related to the van
Vleck determinant. In fact, Eqs.\ (\ref{theta}) and (\ref{Delta2}) imply
\begin{equation}
\left(\frac{\partial q_0}{\partial q_t}\right)_\Theta 
=-\frac{(\partial\Theta/\partial q_t)_{q_0}}
{(\partial\Theta/\partial q_0)_{q_t}}
=\frac{1}{2}\,v(q_0,q_t) 
\left(\frac{\partial \Theta}{\partial q_t}\right)_{q_0}
=-\frac{U'(q_t)\,\Delta}{4\pi g\, v(q_0,q_t)}.
\end{equation}
Eq.\ (\ref{z2.1}) then becomes
\begin{equation}
Z_2(\Theta)= -\frac{1}{4\pi g}\int_{q_{\Theta}^{-}}^{q_{\Theta}^{+}} 
dq_t\,\frac{U'(q_t)\,\Delta^{1/2}}{v(q_0,q_t)}\, e^{-I[q_c]/g}\equiv
\int_{q_{\Theta}^{-}}^{q_{\Theta}^{+}} dq_t\,D(q_t,\Theta)\,e^{-I[q_c]/g},
\label{z2.2}
\end{equation}
where $q_{\Theta}^{\pm}\equiv\lim_{q_0\to\pm\infty}q_t(q_0,\Theta)$. 

The expression above is valid for single-well potentials in general.
Now, let us especialize to the potential (\ref{ux}). $I[q_c]$ is
given by (\ref{iqc}), and using (\ref{vanvleck}) and (\ref{qtheta})
one can write $D(q_t,\Theta)$ as
\begin{eqnarray}
D(q_t,\Theta)&=&\frac{(1+q_t^2)^{1/4}}{\sqrt{4\pi g}}
\left[\frac{1-k^2}{k^2}\,\uT
+\frac{2k^2-1}{k^2}\,E(\vT,k)\right.
\nonumber \\
& &+\left.\frac{\cn\,\uT\,
\dn\,\uT}{\sn\,\uT}+(1-k^2)\,\frac{\cn\,\uT\,
\sn\,\uT}{\dn\,\uT}\right]^{1/2}.
\label{Dqt}
\end{eqnarray}
{}From (\ref{q0-qt}) it follows that $q_0\to\infty$ when 
$\cn(\uT,k)=0$, which occurs when $\uT={\rm K}(k)$, where
${\rm K}(k)$ is the Complete Elliptic Integral of the First Kind.
Using (\ref{gqt}), this condition can be written as an equation in $k$:
\begin{equation}
\frac{\Theta}{2\sqrt{2k^2-1}}={\rm K}(k).
\label{utheta}
\end{equation}
The graph of $f(k)\equiv 2\sqrt{2k^2-1}\,{\rm K}(k)$ is plotted in
Fig.\ \ref{ktheta}. It increases monotonically from zero (at $k=1/\sqrt{2}$)
to infinity (as $k\to 1$), and so for each nonnegative value of $\Theta$
Eq.\ (\ref{utheta}) has a unique solution, which we denote
by $k_{\Theta}$. 
Eq.\ (\ref{gqt}) then gives the corresponding value of $q_{\Theta}^{+}$
($q_{\Theta}^{-}=-q_{\Theta}^{+}$, since $U(-q)=U(q)$).


\subsubsection{Limiting cases of the quadratic approximation}

Expression (\ref{z2.2}) may be used to compute $Z_2(\Theta)$
numerically for 
any value of $\Theta$. However, certain limiting cases may be dealt with 
analytically. These limits are: the harmonic oscillator ($g\to 0$), high 
temperatures ($\Theta\to 0$), and low temperatures ($\Theta\to\infty$).

The limit $g\to 0$ of (\ref{z2.2}) does yield the partition function of the 
harmonic oscillator, as required, 
since $V(x)=\frac{1}{2}\,m\omega^2x^2$ when $g=0$.
In order to arrive at this result, we note that
in this limit one can perform the integral (\ref{z2.2}) using the steepest 
descent method. Details of the derivation can be found in \cite{annals}. 

At high temperatures, $\Theta\to 0$ and (\ref{utheta}) is solved for 
$k_\Theta\to 1/\sqrt{2}$, and so $q_{\Theta}^{+}\to\infty$. It follows that
\begin{equation}
Z_2(\Theta)\stackrel{\Theta\to 0}{\sim}
\sqrt{\frac{1}{2\pi g\Theta}}\int_{-\infty}^\infty dq\, e^{-\Theta\,U(q)/g},
\label{z2theta0}
\end{equation}
or, equivalently,
\begin{equation}
Z_2(\beta)\stackrel{\beta\to 0}{\sim}
\sqrt{\frac{m}{2\pi\hbar^2\beta}}\int_{-\infty}^{\infty}dx\,e^{-\beta V(x)},
\end{equation}
with $V(x)$ and $U(q)$ defined in (\ref{vx}) and (\ref{ux}). This is, 
clearly, the ``classical'' limit for the partition function with a 
pre-factor that incorporates quantum fluctuations.

At low temperatures, $\Theta\to\infty$ and (\ref{utheta}) is 
solved for $k_\Theta\to 1$. A careful derivation \cite{annals} leads to
\begin{equation}
Z_2(\Theta)\stackrel{\Theta\to\infty}{\sim}
\int_{-q_{\Theta}^{+}}^{q_{\Theta}^{+}}\frac{dq_t}{\sqrt{4\pi g}}\,
e^{-I[q_c]/g},
\label{Z2.1}
\end{equation}
with $q_{\Theta}^{+}$ given by
\begin{equation}
q_{\Theta}^{+}=\sqrt{\frac{2{k'}_{\Theta}^2}{1-2{k'}_{\Theta}^2}}
\approx 4\sqrt{2}\,e^{-\Theta/2},
\label{qT+}
\end{equation}
and $I[q_c]$ given by  
\begin{equation}
I[q_c]=\frac{4}{3}\left[\left(1+\frac{1}{2}\,q_t^2\,\nc^2\uT\right)^{3/2}
-1\right]+{\cal O}\left(\Theta\,e^{-\Theta}\right).
\label{ILT}
\end{equation}


\subsubsection{Applications}
\label{applic}

We shall now apply the quadratic semiclassical approximation to obtain the 
ground-state energy and the curve for the specific heat as a function of 
temperature. These two applications will teach us about the power of 
the approximation.
 
In order to compare (\ref{Z2.1}) with the expected low-temperature
limit of the partition function, 
$Z(\Theta)\sim e^{-\Theta\,\varepsilon_0(g)}$ 
(where $\varepsilon_0(g)\equiv E_0(g)/\hbar\omega$ is the dimensionless
ground state energy), it is convenient to rewrite it in a form in
which the $\Theta$-dependence can be analyzed more easily.
This can be done by changing the integration variable back to $q_0$.
Since $q_t\,\nc\,\uT=q_0$ and $q_{\Theta}^{+}$ is the value of $q_t$
corresponding to $q_0\to\infty$, one has
\begin{equation}
Z_2(\Theta)\stackrel{\Theta\to\infty}{\sim}
\int_{-\infty}^{\infty}\frac{dq_0}{\sqrt{4\pi g}}\left(
\frac{\partial q_t}{\partial q_0}\right)_{\Theta}\,
\exp\left\{-\frac{4}{3g}\left[\left(1+\frac{1}{2}\,q_0^2\right)^{3/2}
-1\right]\right\}.
\label{Z2.2}
\end{equation}
When $\Theta\gg 1$ it is possible to write an approximate 
expression for $q_t(q_0,\Theta)$ \cite{annals}, thus allowing to 
write the integrand in (\ref{Z2.2}) solely in
terms of $q_0$ and $\Theta$. The final result is
\begin{equation}
Z_2(\Theta)\stackrel{\Theta\to\infty}{\sim}\frac{2\,e^{-\Theta/2}}
{\sqrt{\pi g}}\int_{-\infty}^{\infty}dq_0\,\frac{
\exp\left\{-\frac{4}{3g}\left[\left(1+\frac{1}{2}\,q_0^2\right)^{3/2}
-1\right]\right\}}{\sqrt{1+\frac{1}{2}\,q_0^2}\left(1+\sqrt{1+\frac{1}{2}\,
q_0^2}\right)}.
\label{Z2.3}
\end{equation}
This gives $\varepsilon_0(g)=1/2$, indicating that the quadratic 
approximation is insufficient to yield corrections to 
the ground state energy of the harmonic oscillator. 
On the other hand, if one recalls that the partition function
can be written as
\begin{equation}
Z(\Theta)=\int_{-\infty}^{\infty}\rho(\Theta;q,q)\,dq,
\label{recall}
\end{equation}
where
\begin{equation}
\rho(\Theta;q,q)=\sum_{n}e^{-\Theta\varepsilon_n}\,|\psi_n(q)|^2
\stackrel{\Theta\to\infty}{\sim} e^{-\Theta\varepsilon_0}\,|\psi_0(q)|^2
\label{rho}
\end{equation}
is the diagonal element of the density matrix, one may take the
square root of the integrand in (\ref{Z2.3}) as an approximation
to the (unnormalized) wave function of the ground state. 
To test the accuracy of this approximation, we have
evaluated the expectation values of the energy for some values
of $g$ and compared them with high precision results found in the
literature. As Table \ref{T1} shows, the ground state energy
computed with this ``semiclassical'' wave function differs from
the exact one by less than $1\%$ even for $g$ as large as 2.

Another concrete problem that can be treated is the calculation of the
specific heat of the quantum anharmonic oscillator. 
It can be written in terms of $Z(\Theta)$ as
\begin{equation}
C=\Theta^2\left[\frac{1}{Z}\,\frac{\partial^2 Z}{\partial\Theta^2}
-\left(\frac{1}{Z}\,\frac{\partial Z}{\partial\Theta}\right)^2\right].
\label{SH}
\end{equation}
This expression was computed using MAPLE for a few values
of $\Theta$ and the coupling constant value $g=0.3$. 
The result is depicted in Fig.\ \ref{specheat},
which also exhibits the curve of specific heat of
the {\em classical} anharmonic oscillator (solid line).
As expected, the results agree when
the temperature is sufficiently high, but, in contrast to
the classical result, the semiclassical approximation
is qualitatively correct at low temperatures too,
dropping to zero as $T\to 0$.

This result, together with the estimate for the ground-state obtained 
previously, shows that the quadratic approximation works very well, being 
quite accurate at high temperatures, and still reliable at lower 
temperatures. In the next subsection, we will comment on why this is so.


\subsubsection{Beyond quadratic}

In this subsection, we shall compute a first correction $G_1$ to the 
quadratic approximation, which corresponds to the $m=1$ term in 
(\ref{zpow}). Using (\ref{deltaU}), we obtain
\begin{equation}
\label{deltaZ}
Z(\Theta)=Z_2(\Theta)-\frac{1}{g}\int_{-\infty}^{\infty}dq_0\,e^{-I[q_c]/g}
\int_0^\Theta d\theta\left[q_c(\theta)
\langle\eta^3(\theta)\rangle+\frac{1}{4}\,\langle\eta^4(\theta)
\rangle\right]+\ldots.
\end{equation}
Eq.\ (\ref{sumP}) yields $\langle\eta^3\rangle=0$ and
$\langle\eta^4\rangle=3g^2\,G_c(0,0;\Theta,0)\,
{\cal G}^2(\theta,\theta)$. Inserting these results in
(\ref{deltaZ}) and changing the integration variable from
$q_0$ to $q_t$ gives
\begin{equation}
Z(\Theta)=\int_{q^-_\Theta}^{q^+_\Theta} dq_t\, D(q_t,\Theta)\,
e^{-I[q_c]/g}\left[1-ga_1(q_t,\Theta)+\ldots\right],
\label{zcorr}
\end{equation}
where
\begin{equation}
a_1(q_t,\Theta)=\frac{3}{4}\int_0^{\Theta}d\theta\,
{\cal G}^2(\theta,\theta).
\label{a1}
\end{equation}
Because of the complicated form of ${\cal G}(\theta,\theta)$, 
it is not a simple task to compute $a_1(q_t,\Theta)$.
However, we can estimate the magnitude of this term
without much effort. Indeed, as shown in \cite{annals},
${\cal G}(\theta,\theta)$ obeys the following inequality:
\begin{equation}
{\cal G}(\theta,\theta)\le\frac{\theta(\Theta-\theta)}{\Theta}
\qquad(0\le\theta\le\Theta).
\label{ineq}
\end{equation}
Therefore,
\begin{equation}
a_1(q_t,\Theta)\le\frac{\Theta^3}{40}.
\end{equation}
This shows that this correction to the quadratic approximation, 
Eq.\ (\ref{z2.1}) or (\ref{z2.2}), can be neglected whenever the condition 
$g\Theta^3/40\ll 1$ is satisfied; this is  compatible with the numerical 
agreements obtained in the applications of the quadratic approximation.

The next term in the expansion for $Z(\Theta)$, which corresponds to the 
$m=2$ term in (\ref{zpow}),  
has a piece with a factor $g$ and one with a factor $g^2$. The former 
comes from the product of $\langle\eta^6\rangle \sim g^3$ with the overall 
$g^{-2}$, whereas the latter involves $\langle\eta^8\rangle \sim g^4$. 
The Feynman diagrams which correspond to the $m=1$ and $m=2$ contributions 
are depicted in Fig.\ \ref{diag1}. Note that two of the $m=2$ diagrams 
involves the three-leg vertex in (\ref{deltaU}), which depends explicitly 
on $q_c(\theta)$. 
Ordinary perturbation theory corresponds to the subset of graphs 
which do not contain the three-leg vertex with the replacement of 
${\cal G}(\theta,\theta')$ by the corresponding (free) expression for the 
harmonic oscillator. In fact, ${\cal G}(\theta,\theta')$ can be expanded 
in terms of its (free)
harmonic oscillator expression and insertions of $(U''[q_c]-1)$, already an 
indication of its non-perturbative nature. Alternatively, we may obtain 
the perturbation theory diagrams by letting $q_c(\theta) \to 0$. 


\section{Conclusions}
\label{conclusions}

The results of section \ref{statistical} can be generalized to 
higher-dimensional Quantum Statistical Mechanics, just as in Quantum 
Mechanics, where this was accomplished in 
\cite{morette1,mizrahi}. The generalization to 
potentials which allow for more than one classical solution, such as the 
double-well quartic anharmonic oscillator, requires a subtle matching of 
the series around each appropriate saddle-point (i.e., the minima). This is 
presently under investigation \cite{joras}.

An extension of our results to field theories is hampered by the fact that 
we do not know how to construct semiclassical propagators in general. The 
technical simplifications which appear in Quantum Mechanics cease to exist. 
However, our methods may still be of use in problems where classical 
solutions have a lot of symmetry (e.g., 
spherical symmetry) so that we can reduce them to effective 
one-dimensional problems. There are many such examples in Physics: 
instantons, monopoles, vortices and solitons are a few of the backgrounds 
that fall into that category. We are currently pursuing this line of 
investigation.

Finally, we should remark that the field-theoretic treatment can be used to 
compute any correlation function of 
interest, in the usual manner. Therefore, a semiclassical series can be 
written down for any physical quantity once it is expressed in terms of 
the relevant correlation functions. 


\acknowledgements

The authors acknowledge support from CNPq, FAPERJ, FAPESP and FUJB/UFRJ.
RMC was also supported in part by the NSF under Grant No.\ PHY94-07194. 
CAAC thanks the organizers of the Workshop for their kind hospitality.
 

\appendix

\section{}
\label{A}

It remains to show how one can obtain
$G_c(\theta_1,\eta_1;\theta_2,\eta_2)$ from the classical path. For this, 
we use the fact that the action $I_2$ is quadratic in $\eta$, and so the 
path integral in (\ref{gc})
is completely determined by the extremum $\eta_e(\theta)$ of 
$I_2[\theta_1,\theta_2;\eta]$, which satisfies Eq.\ (\ref{extremum}), 
subject to the boundary conditions $\eta(\theta_1)=\eta_1$ and 
$\eta(\theta_2)=\eta_2$. Thus,
\begin{equation}
G_c(\theta_1,\eta_1;\theta_2,\eta_2)=G_c(\theta_1,0;\theta_2,0)\,
e^{-I_2[\theta_1,\theta_2;\eta_e]/g},
\label{gc2}
\end{equation}
where, after an integration by parts, 
\begin{equation}
I_2[\theta_1,\theta_2;\eta_e]=\frac{1}{2}\left[\eta_2\,\dot\eta_e(\theta_2)
-\eta_1\,\dot\eta_e(\theta_1)\right].
\end{equation}

We can obtain $\eta_e(\theta)$ by finding the linear combination of any two 
linearly independent solutions, $\eta_a(\theta)$ and $\eta_b(\theta)$, of 
(\ref{extremum}) which satisfies $\eta_e(\theta_1)=\eta_1$ and 
$\eta_e(\theta_2)=\eta_2$. The result is
\begin{equation}
\eta_e(\theta)=\frac{\eta_1\,\Omega(\theta,\theta_2)
+\eta_2\,\Omega(\theta_1,\theta)}{\Omega(\theta_1,\theta_2)},
\end{equation}
with $\Omega(\theta,\theta')$ as defined in (\ref{omega}).
We may then write
\begin{equation}
I_2[\theta_1,\theta_2;\eta_e]=\frac{1}{2\,\Omega_{12}}\,[W_{12}\,\eta_2^2
+W_{21}\,\eta_1^2-(W_{11}+W_{22})\,\eta_1\eta_2],
\label{i2'}
\end{equation}
where $\Omega_{ij}\equiv\Omega(\theta_i,\theta_j)$ and
$W_{ij}\equiv\partial\Omega_{ij}/\partial\theta_j$.
(Note that $W_{ii}$ is the Wronskian of $\eta_a$ and 
$\eta_b$ computed at $\theta_i$.)

Explicit expressions for $\eta_a(\theta)$ and $\eta_b(\theta)$
can be obtained as follows.
By differentiating (\ref{euler}) with respect to $\theta$, one can verify 
that $\eta_a(\theta)=\dot q_c(\theta)$ satisfies (\ref{extremum}).
For the second solution, we take
$\eta_b(\theta)=\dot q_c(\theta)\,Q(\theta)$,
where $Q(\theta)$ is defined as
\begin{equation}
Q(\theta)=Q(0)+\int_0^\theta\frac{d\theta'}{{\dot q}^2_c(\theta')}
\label{qthetaq0}
\end{equation}
for $\theta<\Theta/2$, $Q(\theta)=-Q(\Theta-\theta)$ for
$\theta>\Theta/2$, and $Q(0)$ is chosen so as to make
$\dot\eta_b(\theta)$ continuous at $\theta=\Theta/2$. One can easily check, 
using (\ref{euler}), that $\eta_b(\theta)$ 
indeed satisfies (\ref{extremum}). (Alternatively, one could use a 
procedure  
introduced by Cauchy \cite{morette1,mizrahi}, and differentiate the 
classical 
solution $q_c(\theta)$ with respect to any two parameters related to
its two constants of integration.) We can now write explicit expressions
for $\Omega_{12}$ and $W_{ij}$:
\begin{equation}
\Omega_{12}=\dot q_c(\theta_1)\,\dot q_c(\theta_2)\,
[Q(\theta_2)-Q(\theta_1)],
\label{o12}
\end{equation}
\begin{equation}
W_{ij}=\dot q_c(\theta_i)\, U'[q_c(\theta_j)]\,[Q(\theta_j)-Q(\theta_i)]
+\frac{\dot q_c(\theta_i)}{\dot q_c(\theta_j)}.
\label{w12}
\end{equation}

As a final step, the pre-factor in (\ref{gc2}) can be derived \cite{annals} 
using the methods of Refs.\ \cite{zinnjustin,wileyboyanholman}. 
The result is 
\begin{equation}
G_c(\theta_1,0;\theta_2,0)={\left[\frac{W_{11}}{2\pi g\,\Omega_{12}}
\right]}^{1/2}.
\end{equation}
{}From (\ref{w12}), one easily finds $W_{ii}=1$. Therefore, our quadratic 
semiclassical propagator is given by 
\begin{equation}
G_c(\theta_1,\eta_1;\theta_2,\eta_2)=\frac{1}{\sqrt{2\pi g\,\Omega_{12}}}\,
\exp\left[-\frac{1}{2g\,\Omega_{12}}\,(W_{12}\,\eta_2^2
+W_{21}\,\eta_1^2-2\,\eta_1\eta_2)\right].
\label{gs}
\end{equation}
As promised, it is completely determined by the classical solution. 

Finally, we note that the van Vleck determinant $\Delta$ is a by-product of 
(\ref{gs}):
\begin{equation}
\Delta(q_0,\Theta)=G_c^{-2}(0,0;\Theta,0)=2\pi g\,\Omega(0,\Theta)
=4\pi g\,\dot{q}_c^2(0)\,Q(0).
\label{vanvleck}
\end{equation}
As shown explicitly in \cite{annals}, one can express $\Delta$ as
\begin{equation}
\Delta=\frac{4\pi g\,[U(q_t)-U(q_0)]}{U'(q_t)}\left(\frac{\partial\Theta}
{\partial q_t}\right)_{q_0}.
\label{Delta2}
\end{equation}
Together with (\ref{action}), this shows that one does not need to know 
$q_c(\theta)$ in order to write the first 
term in the semiclassical series; it is enough to know
$q_t(q_0,\Theta)$.



\begin{table}

\caption{Ground state energies for different values of $g$
($\hbar=m=\omega=1$).}
\label{T1}

\begin{tabular}{c c c c}
$g$ & $E_0$(semiclassical)\tablenote[1]{$\langle\phi_0|H|\phi_0\rangle/
\langle\phi_0|\phi_0\rangle$, where $\phi_0(q_0)$ is the square root
of the integrand in Eq.\ (\ref{Z2.3}).} 
& $E_0$(exact)\tablenote[2]{Values quoted from
Ref.~\cite{vinette}.} & error($\%$) \\
\hline
0.4 & 0.559258 & 0.559146 & 0.02 \\
1.2 & 0.639765 & 0.637992 & 0.28 \\
2.0 & 0.701429 & 0.696176 & 0.75 \\
4.0 & 0.823078 & 0.803771 & 2.40 \\
8.0 & 1.011928 & 0.951568 & 6.34 \\
\end{tabular}

\end{table}


\begin{figure}
\caption{$U(q)$.}
\label{potential}
\end{figure}

\begin{figure}
\caption{Graph of $f(k)$.}
\label{ktheta}
\end{figure}

\begin{figure}
\caption{Specific heat vs.\ temperature ($T=1/\Theta$) for the 
quantum (diamonds) and classical (solid line) anharmonic
oscillator. $g=0.3$.}
\label{specheat}
\end{figure}

\begin{figure}
\caption{Feynman graphs for $m=1$ and $m=2$.}
\label{diag1}
\end{figure}


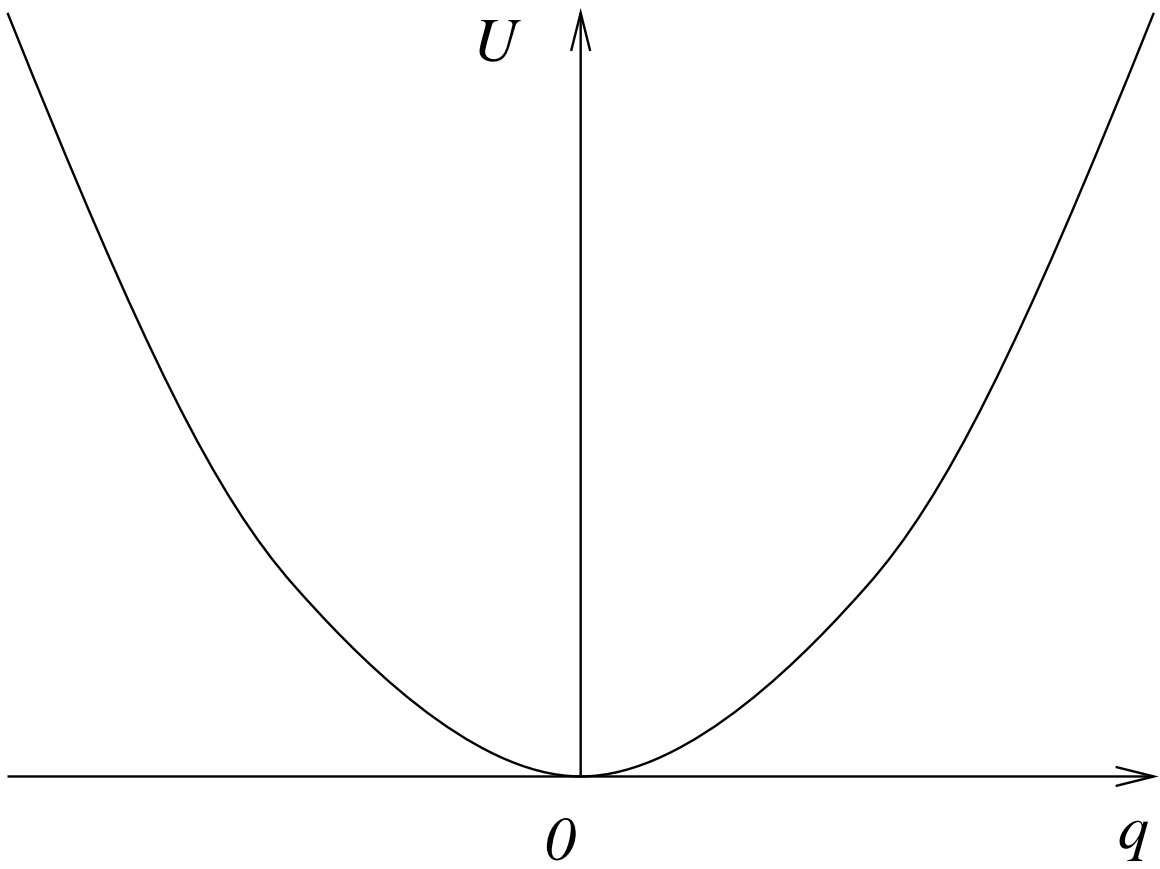
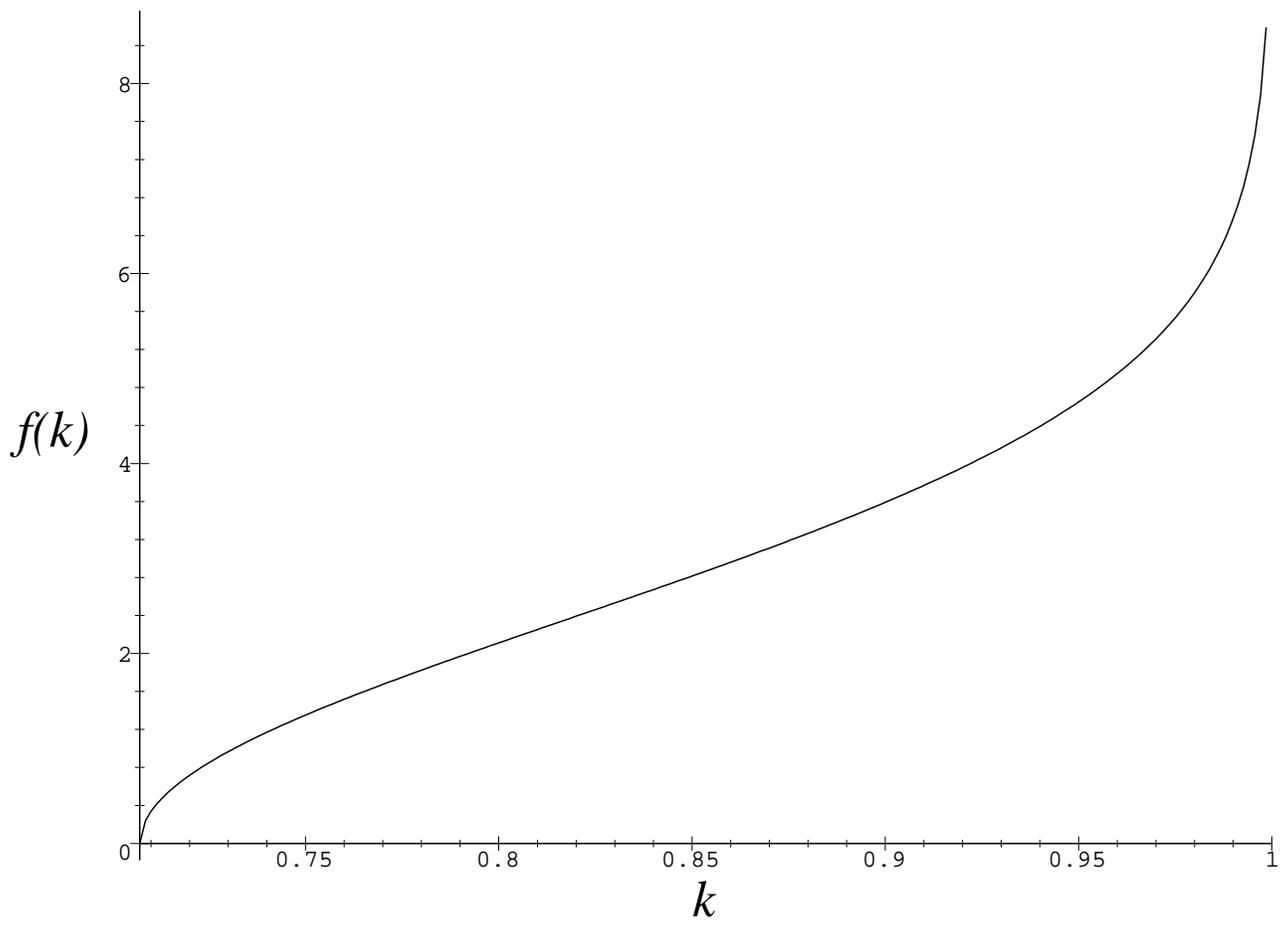
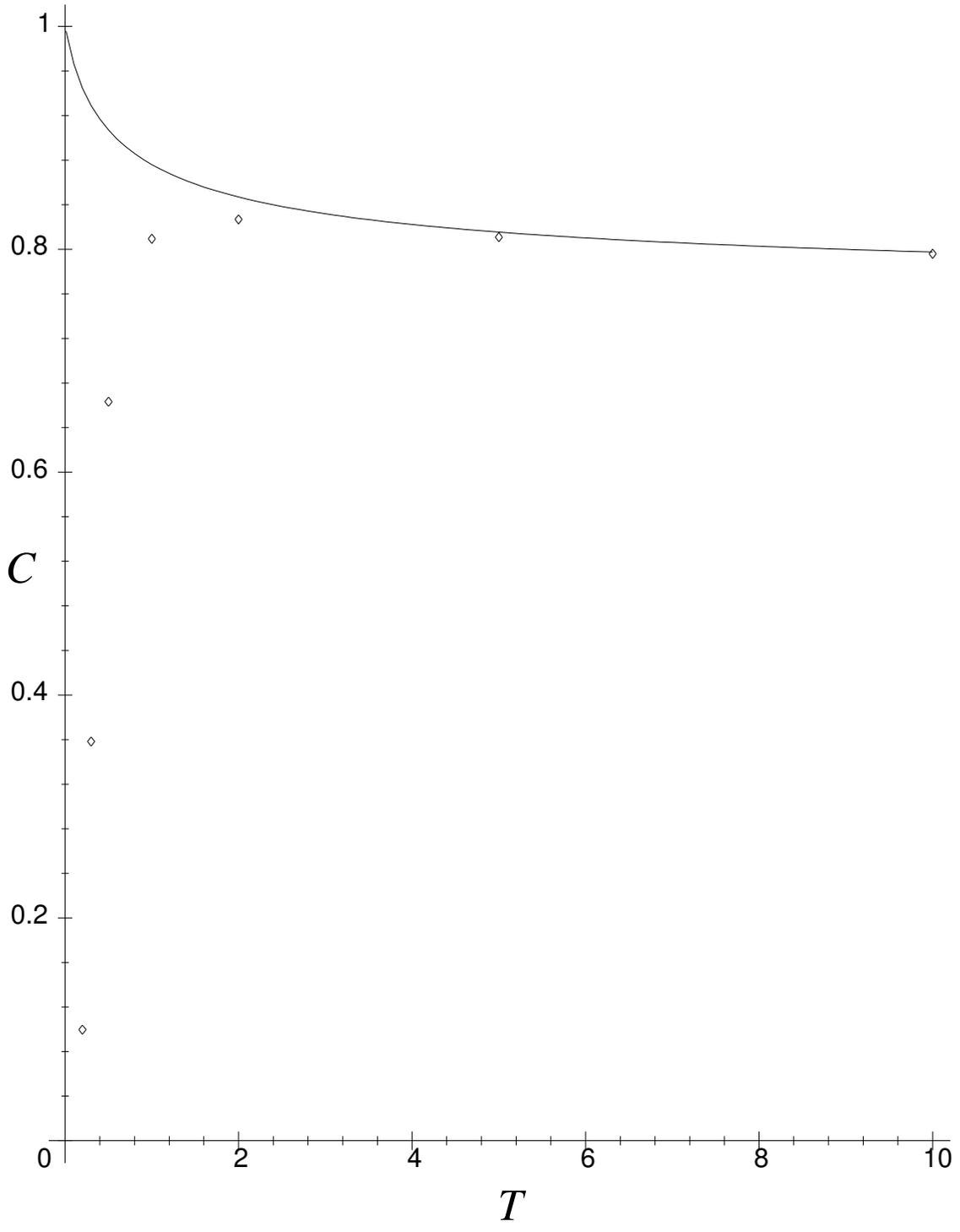
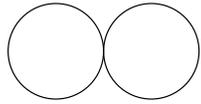
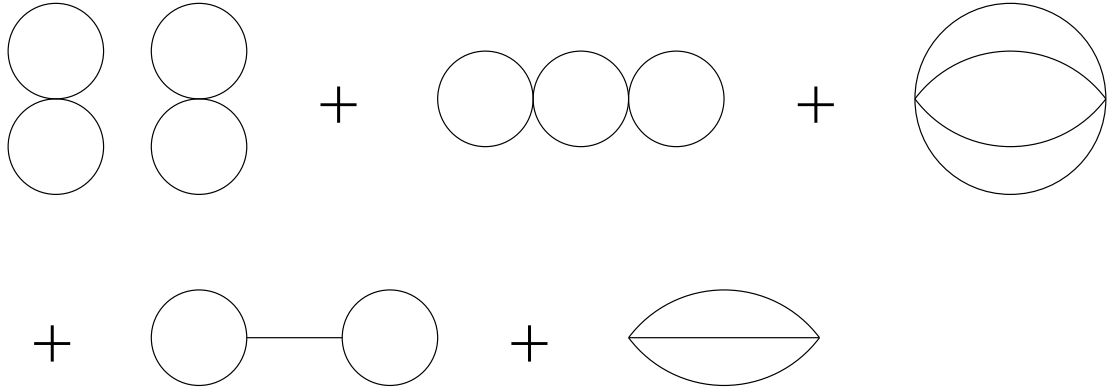
\end{document}